\documentclass[prl,aps,amsfonts,amssymb,floats,
twocolumn,showpacs]{revtex4}
\usepackage{epsf}
\usepackage{graphicx}

\begin{document}

\title{On the critical level-curvature distribution}
 
\author{S.N. Evangelou\footnote{e-mail: sevagel@cc.uoi.gr,
permanent address: Physics Department, University of Ioannina,
Ioannina 45110, Greece}}

\affiliation{Max-Planck-Institut for the Physics of Complex Systems,
Noethnitzer Strasse 38, D-01187 Dresden, Germany}

 
\begin{abstract} 
The parametric motion of energy levels for
non-interacting electrons at the Anderson localization 
critical point is studied by computing the energy level-curvatures 
for a quasiperiodic ring with twisted boundary conditions.
We find a critical distribution which has the universal 
random matrix theory form ${\bar P}(K)\sim |K|^{-3}$
for large level-curvatures $|K|$ corresponding to quantum diffusion,
although overall it is close to approximate log-normal statistics 
corresponding to localization. The obtained hybrid distribution 
resembles the critical distribution of the disordered Anderson model 
and makes a connection to recent experimental data.
\end{abstract} 
 

\maketitle 
\narrowtext 
 

\par
\medskip
The Landauer-Buttiker(LB) scattering approach\cite{1} nowadays
is almost exclusively used for computing the conductance $g$
for many materials in nanoscience. 
This very appealing formalism although not extremely rigorous
is particularly suitable for small systems, 
something which became obvious following 
the discovery of conductance steps in ballistic 
point contacts more than a decade ago \cite{2}.
In such a scattering approach one gets the non-equilibrium
transport for any suffiently realistic system 
if it is connected to two ideal leads. The
conductance is simply the transmission  probability 
through the sample scatterer  which can be 
evaluated at any energy monitored by gates, 
e.g. via Green's function  techniques.
Although $g$ is strictly speaking 
a property of open systems 
it can be also probed for a closed system 
from the eigensolutions of the stationary 
Schroedinger's equation.
This is done via Thouless's intuitive definition of
conductance which involves a measure of the sensitivity 
of eigenvalues to twisted boundary conditions (BC) \cite{3}.
In the later approach, the conductance is related
to the so-called level-curvatures  and is usually expressed
by the ratio of the geometric mean of the absolute
curvature at zero twist over the local mean 
level-spacing.
The scattering approach for open systems and the 
eigenvalue shift approach for closed systems are  
two established definitions of conductance, which 
compare rather favorably with each other for electrons 
in disordered systems (see Ref. \cite{4}).

\par
\medskip
The search for universal features in the
statistics of stationary electronic levels 
and  their motion as a function 
of some parameter is a large area of study
with  obvious consequences to quantum transport.
For example, avoided level-crossings which characterize
the spectrum of a chaotic system  under a perturbation, 
such as the application of external fields,     
evolve with universal random matrix
theory (RMT) laws \cite{5}. The very small level-spacings for 
avoided level-crossings correspond to large absolute level-curvatures
so that the tail of the level-curvature distribution ${\bar P}(K)$
is intimately related to the small part of the level-spacing
distribution.
Therefore, in  the absence of symmetry breaking mechanisms
the linear part of the Wigner distribution for small level-spacings  
implies large level-curvatures having a decreasing 
asymptotic behavior ${\bar P}(K)\sim 1/|K|^{2+\beta}$,
with $\beta=1$ \cite{6,7,8}.  This has been confirmed
experimentally for level-curvature distributions of
acoustic resonance spectra from quartz blocks \cite{9}.

\par
\medskip
We focuse on the critical distribution of level-curvatures
by obtaining the response of a quasiperiodic complex system's
energy spectrum to different BC \cite{10}, within Thouless's 
approach to quantum transport.
For this purpose a phase factor $\exp({\it i}\phi)$ 
is imposed to the hopping probability connecting
the first and the last sites of a ring
and the resulting parametric dependence of the 
energy levels to changes of $\phi$ is obtained.
This allows to explore the nature of the 
corresponding eigenstates from extended to localized,
since extended states simply feel any changes in BC having
large curvatures while localized states are insensitive 
to BC having curvatures which approach zero. 
Nevertheless, the studied model is interesting only
at the metal-insulator transition point since it satisfies
duality for the extended (quasiballistic) 
and the localized (nonrandom) regimes \cite{11}. 
One may wonder what the level-curvature distribution might
be at the metal-insulator transition for such a quasiperiodic 
system which has Cantor set-like fractal electronic structure 
and a semi-Poisson hybrid distribution of level-spacings \cite{10}.

\par
\medskip
Deviations around the  maximum peak between
level-curvatures obtained in experiment \cite{9}
and RMT  were attributed to the presence of 
hidden symmetries \cite{12,13}. 
In our study the critical distribution  
of level-curvatures is obtained in a 
convenient one-dimensional setting which allows 
to discuss both the three-dimensional disordered system 
and questions of universality in general.
Our finding is a scale-invariant ${\bar P}(K)$
which resembles the distribution obtained for the 
normalized level-curvatures in $3D$ critical
disordered systems \cite{14,15},
having both a diffusive tail and overall localized behavior \cite{4}. 
For a finite system, the transition from extended 
to localized states involves broadening of the distribution  
${\bar P}(K)$ and lowering its maximum peak as the 
curvatures move to lower values by increasing disorder.
Our study, on one one hand, is in agreement with the semi-Poisson 
level-spacing distribution for the same system \cite{10}. 
On the other hand, the coexistence of the localized  
almost log-normal broad form with the diffusive tail 
might be the reason for the lowering of the maximum peak   
observed in the experiment \cite{9}.


\par
\medskip
We have studied the distribution function 
of level-curvatures, defined as  
\begin{equation}
K_{\alpha}= \left. {\frac
{d^{2}\varepsilon_{\alpha}}{{d\phi}^{2}}} \right\vert_{\phi=0},
\end{equation}
for the energy levels $\varepsilon_{\alpha}$  of 
a quasiperiodic ring in the presence of phase 
$\exp({\it i}\phi)$, obtained from 
the critical tight-binding model Hamiltonian \cite{10}
\begin{eqnarray}
H=\sum_{n}V_{n}c^{\dag }_nc_m- \sum_{\langle nm \rangle }(c^{\dag
}_nc_m+c^{\dag }_mc_n)\nonumber \\
-\sum_{\langle nm \rangle
}\left(\exp({{\it i} \phi})c^{\dag }_nc_m+\exp({-{\it i}
\phi})c^{\dag }_mc_n
\right).
\end{eqnarray}
The sum is taken over all sites $n$,
$c_n$ ($c_{n}^{+}$) is the annihilation (creation)
operator on site $n$ and the potential 
$V_{n}=2cos(2\pi\sigma n)$ is chosen at criticality 
with $\sigma$ the golden mean irrational.
The eigenvalues $\varepsilon_{\alpha}$ have corresponding
eigenvectors $|\alpha\rangle=\sum_{x=1,L} \psi_{\alpha}(x)|x\rangle$ 
with amplitudes $\psi_{\alpha}(x)=\langle x|\alpha\rangle$ 
for a finite chain of size $L=F_{i}$ with 
$\sigma=F_{i-1}/F_{i}$,  
the ratio of two successive Fibonacci numbers $F_{i-1}$, 
$F_{i}$.  The imposed general boundary condition
to the wave function $\psi(x+L)=e^{{\it i}\phi}\psi(x)$
is equivalent to piercing the ring by the Aharonov-Bohm 
magnetic flux $\Phi$ via $\phi=2\pi\Phi/\Phi_{0}$, with  
$\Phi_{0}=c\hbar/e$ the flux quantum.

\par
\medskip
The curvatures $K_{\alpha}$ are properties of the energy 
levels in the limit of zero flux so that perturbation theory 
can be applied \cite{4}. Using the small-$\phi$ expansion 
$exp(\pm {\it i}\phi)=1\pm {\it i}\phi-\phi^{2}/2+...$
the relevant terms up to $2$nd order 
are $<\alpha|V|\alpha>= -\phi^{2} <\alpha|1> <1|\alpha>$ and 
$<\alpha|V|\beta>= {\it i} \phi( <\alpha|1> <N|\beta>-
 <\alpha|N> <1|\beta>)$ so that
the level-curvature for the eigenvalues $\varepsilon_{\alpha}$ is
\begin{equation}
K_{\alpha}=2\psi_{\alpha}(L)\psi_{\alpha}(1)
+ 2\sum_{\beta\neq\alpha }{\frac
{\left(\psi_{\beta}(L)\psi_{\alpha}(1)-\psi_{\beta}(1)\psi_{\alpha}(L)
\right)^{2}}{\varepsilon_{\beta}-\varepsilon_{\alpha}}} \nonumber \\
\end{equation}
$\alpha=1,2,...,L,$
in terms of the eigenvalues $\varepsilon_{\alpha}$
and eigenvector amplitudes $\psi_{\alpha}$ of the unperturbed
ring ($\phi=0$).
The level-curvatures  $K_{\alpha}$ give quantum transport
properties and they are more difficult to obtain than level-spacings
since one also needs the eigenvectors of the Hamiltonian.
The formula of Eq. (3) is exact as long as $\varepsilon_{\beta}\ne
\varepsilon_{\alpha}$, since higher orders vanish at $\phi=0$.
Moreover, it is also both conceptually and numerically
meaningful since the wave function amplitudes 
at the two ends of the chain  $\psi_{\alpha}(1)$, $\psi_{\alpha}(L)$ 
are expected to play a role for transport,
while linear algebra relates the first term in Eq. (3)
to the error in eigenvalue determination from truncating 
an infinite tridiagonal matrix  \cite{16}.
 
\par
\medskip
To use Eq. (3) we only need the eigensolutions
of the periodic BC problem, that is the solutions
of the Hamiltonian of Eq. (2) with $\phi=0$. 
The problem simplifies further by the presence of the parity 
symmetric potential $cos(x)=cos(-x)$  which allows to find
symmetric and antisymmetric solutions separately, 
dealing  with two tridiagonal matrices instead of the full 
matrix in the presence of BC \cite{17}.
This dramatic reduction in storage requirements
is achieved if the one-electron basis is rearranged to run
within $-s, -s+1, ..., s-1, s$ also by
distinguishing between even and odd size $L$. 
For example, for odd $L=2s+1$ the problem is
reduced to a tridiagonal matrix of size $s+1$
for the symmetric states which has diagonal matrix elements
the potential values $V_{0}, V_{1}, ...,V_{s-1}, V_{s}+1$, unity
elements lying next to the diagonal, except in
the first (second) row second(first) column where 
is $\sqrt{2}$, and each eigenvector amplitude 
$(\chi(0),\chi(1),...,\chi(s))$ is 
related to the original via the symmetric rule
$\psi(0)=\chi(0)$, $\psi(-s)=\psi(s)=\chi(s)/\sqrt(2)$.
The second tridiagonal matrix for the antisymmetric states 
is of size $s$ with matrix elements $V_{1}, ...,V_{s-1}, 
V_{s}-1$ on the diagonal and unity next to diagonal with
corresponding eigenvector amplitudes $(\chi(1),...,\chi(s))$
related to the original via antisymmetry $\psi(0)=0$, 
$\psi(-s)=-\psi(s) =-\chi(s)/\sqrt(2)$.
By replacing $\psi$ with $\chi$ Eq. (3) 
splits into a symmetric part labelled by $\alpha=1,2,...,s+1$
\begin{eqnarray}
K_{\alpha}=(\chi_{\alpha}(s))^{2} + 2\sum_{
\beta=s+1}^{2s+1}{\frac
{\left(\chi_{\beta}(s)\chi_{\alpha}(s)
\right)^{2}}{\varepsilon_{\beta}-\varepsilon_{\alpha}}},
\end{eqnarray}
and an antisymmetric part for $\alpha=s+2,s+3,...,2s+1$
\begin{eqnarray}
K_{\alpha}=-(\chi_{\alpha}(s))^{2} + 2\sum_{
\beta=1}^{s+1}{\frac
{\left(\chi_{\beta}(s)\chi_{\alpha}(s)
\right)^{2}}{\varepsilon_{\beta}-\varepsilon_{\alpha}}}.
\end{eqnarray}
In Eqs. (4), (5) the $s$-th last element of each eigenvector  
can be computed iteratively  with no need to increase the storage 
requirements beyond that of a tridiagonal matrix. Moreover, the
corresponding sums over $\beta$ 
run over opposite kind of symmetric and antisymmetric states,
respectively, being precisely zero for 
species of the same symmetry.  
A similar formula for even $L$ is easily obtained.


\begin{figure}
\includegraphics[scale=0.4]{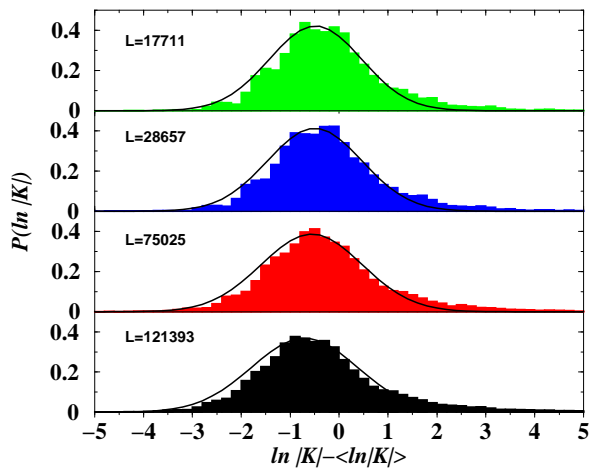}
\caption{The computed critical  distribution for  the
logarithm of level-curvatures $P(\ln |K|)$ and various sizes $L$
is shown to be almost scale-invariant. 
The histograms for each size are obtained from the diagonalization 
of Eq. (2) employing the formula for the level-curvatures of Eqs. (4),(5).
The continuous line is a Gaussian fit of the data.}
\end{figure}


\par
\medskip
A  ballistic ring has
$K_{\alpha}=(8\pi^{2}/L^2)\cos(2\pi\alpha/L)$
which give square-root singularities for ${\bar P}(K)$.
These disappear for the critical ring studied where 
the curvature distribution ${\bar P}(K)$ is much more complex 
so it turns out more convenient to consider the logarithmic 
distribution $P(\ln |K|)=|K|{\bar P}(K)$, instead.
The computed critical distribution
presented in Fig. 1 for various sizes $L$ 
is shown to be scale-invariant,
overall approximated by a logarithmic normal 
form. In Fig. 2 we present  a log-log 
plot of $P(\ln |K|)$  which demonstrates the 
diffusive universal tail ${\bar P}(K)\sim |K|^{-3}$
for large-$|K|$.
For small-$|K|$ we approximately find
${\bar P}(\ln|K|) \sim |K|^{2}$.
Unfortunately, rapid loss of accuracy 
for the too small curvatures 
does not permit to extract reliable exponents 
from fits of ${\bar P}(K)$ to the form suggested in \cite{14}.  
However, the overall behavior of the critical distribution 
shown in Fig. 3 is roughly similar to the critical distribution
obtained for the $3D$ Anderson model \cite{14,15}.

\par
\medskip
\begin{figure}
\includegraphics[scale=0.4]{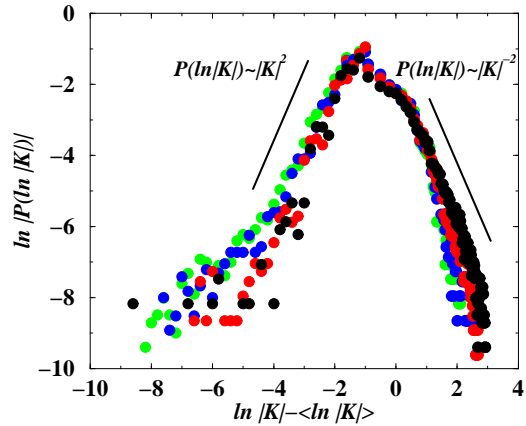}
\caption{Log-log plot of the critical level-curvature
distribution for various sizes $L$ as in Fig. 1.
Apart from the less accurate points in the left corner 
we see a slower behavior  for small-$|K|$ by plotting the 
equivalent of ${\bar P}(K)\sim |K|$, while 
for large-$|K|$ the diffusive asymptotic tail becomes
${\bar P}(K)\sim |K|^{-3}$.}
\end{figure}
\par
\medskip
\begin{figure}
\includegraphics[scale=0.4]{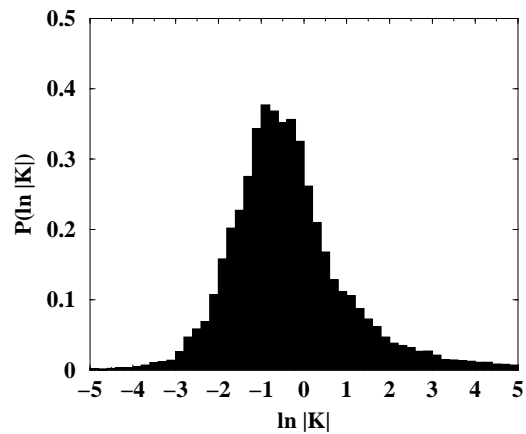}
\caption{Another demonstration of 
the critical level-curvature distribution 
for two sizes $L=17711, 28657$ is shown which resembles
the distribution for a $3D$ disordered system \cite{14,15}. 
The distributions for the quasiperiodic and the disordered system
are rather similar although in the disordered case the curvatures 
are normalized divided by the local mean level-spacing $\Delta$.}
\end{figure}


\par
\medskip
The level-curvatures in disordered or chaotic systems
exploit a sort of ``dynamics" of the quantum 
stationary spectra as a function
of flux which might be thought of  ``time".
The conductance is then obtained from the level-curvatures 
in terms of eigenvalues and eigenvectors 
of the tight binding system with periodic BC.
In our case the computational effort is minimized 
because the parity symmetry of the potential reduces the problem 
to two simple tridiagonal matrices for 
the determination of symmetric and antisymmetric states.
The needed last element of each eigenvector  
can be also computed efficiently with no increase of storage
so that the size of the matrices can easily exceed $10^{5}$.
We remind the reader that for diffusive disordered systems 
the computed averaged LB conductance $\langle g\rangle$ was shown  \cite{4}
to be related to the mean absolute curvature 
via $\langle g\rangle=\pi\langle |K|\rangle/\Delta$, 
where $\Delta$ is the local mean-level spacing. 
For localized disordered systems the curvatures diminish and
both distributions approach a log-normal
form with $\langle \ln g\rangle=\pi\langle \ln(|K|)\rangle$.
Our study of the unormalized level-curvatures 
for the quasiperiodic system enabled us  to obtain
the critical ${\bar P}(K)$ for very large system sizes.  
The curvatures become 
very small for such large systems which set limits to the 
accuracy of the distribution.
Our results are similar to that obtained in $3D$ critical 
disordered system. Moreover, such
computations might suggest that the lowering of the 
maximum peak observed in experiment \cite {9} 
could be thought of as due to an approach 
towards the critical region. Although
the critical distribution shown in Fig. 3 
still displays a diffusive tail of the log-normal form  
it is much broader having less height than the pure 
diffusive RMT one.

\par
\medskip
The obtained results for non-interacting fermions 
in a quasiperiodic ring  complement and confirm previous studies
at the metal-insulator transition of disordered systems. 
The distribution ${\bar P}(K)$ for the ensemble of 
critical states at the transition 
(there is no ensemble over disorder) is scale-invariant 
like the corresponding level-spacing distribution which
is known to depend sensitively  on  BC \cite{10}. 
After average over BC the critical distribution of the 
level-spacings can be described by the semi-Poisson curve which combines 
both extended and localized behavior. Similarly, the obtained 
${\bar P}(K)$  exhibits a hybrid character like
what is obtained for critical $3D$ disordered systems.
In closing, it is remarkable that such a simple
one-dimensional non-random model can capture most 
features displayed at the Anderson metal-insulator 
transition of realistic disordered systems.



\begin{thebibliography}{99}
 
\bibitem{1}  R. Landauer, Phil. Mag. {\bf 21}, 863 (1970);
M. Buttiker, Y. Imry, R. Landauer and S. Pinhas, 
Phys. Rev. {\bf 31}, 6207 (1985); M. Buttiker, Phys. Rev. B {\bf 38},
9375 (1988).

\bibitem{2}  Y. Imry, {\it ``Introduction to Mesoscopic Physics"}, 
Oxford University Press, 1997.


\bibitem{3} J.T. Edwards and D.J. Thouless, J.Phys. C {\bf 5},
807 (1972); D.J. Thouless, Phys. Rep. {\bf 13}, 93 (1974).

\bibitem{4}D. Braun, E. Hofstetter, G. Montambaux and A. MacKinnon,
cond-mat/9611059.

\bibitem{5} T. Guhr, A. M$\ddot{u}$ller-Groeling, and H.
Weidenm$\ddot{u}$ller, Phys. Rep. {\bf 229}, 189-425 (1998).

\bibitem{6}  J. Zakrzewski and D. Delande, Phys Rev. E
 {\bf 47}, 1650 (1993).

\bibitem{7}  F. von Oppen, Phys Rev. E
 {\bf 51}, 2647 (1995).

\bibitem{8}  Y.V. Fyodorov and H.-J. Sommers, Phys Rev. E
 {\bf 51}, R2719 (1995).

\bibitem{9} P. Bertelsen, C. Ellegaard, T. Guhr,
M. Oxborrow and K. Schaadt, Phys. Rev. Lett. {\bf 83}, 2171 (1999).

\bibitem{10}  S.N. Evangelou and J.-L. Pichard,
 Phys. Rev. Lett. {\bf 84}, 1643 (2000).

\bibitem{11}  C. Aulbach, A. Wobst, G.-L. Ingold, P. Hanggi 
  and I. Varga, New. J. Phys. (to appear).

\bibitem{12}  G. Erg${\ddot u}$n and Y.V. Fyodorov,
 Phys. Rev. E {\bf 68}, 046124 (2003).

\bibitem{13}  M.S.  Hussein, C.P. Malta, M.P. Pato and A,P.B. Tufaile,
 Phys. Rev. E {\bf 65}, 057203 (2002).

\bibitem{14} C.M. Canali, C. Basu, W. Stephan and V.E. Kravtsov,
Phys. Rev. B {\bf 54}, 3 (1996) generalized the RMT form in the absence of 
symmetry breaking fields ($\beta=1$) to
${\bar P}(K)\sim (1+K^{\mu})^{-3/\mu}$, where $\mu =2$
for RMT and $\mu\approx 1.6$ for the $3D$ critical disordered system.

\bibitem{15} I.Kh. Zharekeshev and B. Kramer,
Physica A {\bf 266}, 450 (1999).

\bibitem{16} A. Fox {\it ``Introduction to Numerical 
Linear Algebra"}, Oxford: Clarendon University Press, 1964.

\bibitem{17}  D.J. Thouless
Phys. Rev. B {\bf 28}, 4272 (1983).


\end{thebibliography}
\end{document}